\documentclass[prl,printnumbers,superscriptaddress,floatfix]{revtex4}

\usepackage{amsfonts}
\usepackage{graphicx}
\usepackage{graphicx,epsfig,latexsym,amssymb}
\usepackage{multirow,amsmath,array,booktabs}
\usepackage{dcolumn}
\usepackage[section]{placeins}
\usepackage{bm}

\begin{document}

\title{Relativistic symmetry in deformed nuclei by similarity
renormalization group}

\author{Jian-You Guo}
\email[E-mail:]{jianyou@ahu.edu.cn}
\affiliation{School of Physics and Material Science, Anhui University, Hefei 230039,
People's Republic of China}

\author{Shou-Wan Chen, Zhong-Ming Niu, Dong-Peng Li, Quan Liu}
\affiliation{School of Physics and Material Science, Anhui University, Hefei 230039,
People's Republic of China}

\begin{abstract}
The similarity renormalization group is used to transform a general Dirac
Hamiltonian into diagonal form. The diagonal Dirac operator consists of the
nonrelativistic term, the spin-orbit term, the dynamical term, and the
relativistic modification of kinetic energy, which are very useful to
explore the symmetries hidden in the Dirac Hamiltonian for any deformed
system. As an example, the relativistic symmetries in an axially deformed
nucleus are investigated by comparing the contributions of every term to the
single particle energies and their correlations with the deformation. The
result shows that the deformation considerably influences the spin-orbit
interaction and dynamical effect, which play a critical role in the relativistic
symmetries and its breaking.
\end{abstract}

\pacs{21.10.Hw,21.10.Pc,03.65.Pm,05.10.Cc}
\maketitle

It is well known that the spin and pseudospin symmetries play a critical
role in the shell structure and its evolution. The introduction of the
spin-orbit potential to the single-particle shell model can well explain the
experimentally observed existence of magic numbers for nuclei close to the
valley of $\beta $ stability~\cite{Haxel49,Mayer49}. To understand the near
degeneracy between two single-particle states with the quantum numbers ($%
n-1,l+2,j=l+3/2$) and ($n,l,j=l+1/2$), the concept of pseudospin symmetry
(PSS) was introduced by defining the ($\tilde{n}=n-1,\tilde{l}=l+1,j=\tilde{l%
}\pm 1/2$) as the pseudospin doublets~\cite{Hecht69,Arima69}. The doublets
persist for deformed nuclei as well~\cite{Bohr82}. The axially deformed
single-particle orbits with the asymptotic Nilsson quantum numbers ($\Omega
=\Lambda +1/2[N,n_{3},\Lambda ]$) and ($\Omega =\Lambda +3/2[N,n_{3},\Lambda
+2]$) are quasidegenerate, and can be viewed as the pseudospin doublets ($%
\Omega =\tilde{\Lambda}\pm 1/2[\tilde{N}=N-1,\tilde{n}_{3}=n_{3},\tilde{%
\Lambda}=\Lambda +1]$). Since the PSS is suggested in atomic nuclei, there
have been comprehensive efforts to understand its origin. In 1997, the PSS
was shown to be a symmetry of the Dirac Hamiltonian~\cite{Ginoc97}. The
pseudo-orbital angular momentum $\tilde{l}$ is nothing but the orbital
angular momentum of the lower component of the Dirac spinor, and the
equality in magnitude but difference in sign of the scalar potential $S$ and
vector potential $V$ was suggested as the exact PSS limit. Soon
afterwards, this condition was generalized to $d(S+V)/dr=0~$\cite{Meng98}.
As there exist no bound nuclei in the PSS limit, much effort is devoted to
the mechanism of pseudospin breaking. By transforming the Dirac equation
into a Schr\"{o}dinger-like one, the influence of every term on the
pseudospin breaking was checked and the dynamical nature of PSS was
suggested in real nuclei~\cite{Alber01,Lisboa10}. In order to better
understand the symmetry of the Dirac Hamiltonian, the spin symmetry in the
anti-nucleon was studied with the same origin as PSS discovered~\cite{Zhou03}%
. Further, the supersymmetric description of PSS was presented for the
spherical and axially deformed nuclei~\cite{Leviatan04,Typel08,Leviatan09}.
In combination with the perturbative theory, the non-perturbative nature of
PSS was indicated \cite{Liang11,Gonoc11}. Moreover, this symmetry was also
checked in the resonant states ~\cite{Guo05,Lu12}. More reviews on the PSS
can be found in the literatures~\cite{Ginoc05PR,Liang13} and the references
therein.

Regardless of these pioneering studies, the origin of PSS has not been fully
understood in the relativistic framework. Recently, we have applied the
similarity renormalization group (SRG) to Dirac Hamiltonian for a spherical
system, and obtained a diagonal Dirac operator, which consists of the
nonrelativistic term, the spin-orbit term, the dynamical term, the
relativistic modification of kinetic energy, and the other term~\cite{Guo121}%
. In the diagonal Dirac operator, every term mentioned above is Hermitian,
and all the defects in the usual decoupling disappear~\cite{Guo121}, which
is very useful to analyze the PSS hidden in the Dirac Hamiltonian. As
pointed out by Liang and etc.~\cite{Liang13}, the work in Ref.~\cite{Guo121}
fills the gap between perturbation calculations and the supersymmetry
descriptions. They have applied the operator under the lowest-order
approximation to research on the origin of PSS and its breaking mechanism by
the supersymmetry quantum mechanics and perturbation theory~\cite{Liang13}.
By including the lowest-order spin-orbit term, they have further
investigated the spin-orbit effect on the PSS breaking~\cite{Liang132}. We
have applied the operator to check the contributions of every term to the
pseudospin splitting and their influences on the PSS in Refs.~\cite%
{Guo122,Guo13}.

Considering that most of the real nuclei are deformed, in this paper we
apply SRG to a general Dirac Hamiltonian and transform it into a
diagonal form while keeps every term Hermitian. Such a study is
significant for not only nuclei. As pointed out in Ref.~\cite{Leviatan09},
the cylindrical geometries are relevant to a number of problems, including
electron channeling in crystals, structure of axially deformed nuclei, and
quark confinement in spheroidal flux tubes. As an example, we check the relativistic symmetry for an axially deformed nucleus by comparing the contributions of every term to the single particle energies and their correlations with the deformation to disclose the origin of the PSS and its breaking mechanism in deformed nuclei.

For simplicity, we sketch our formalism with the following Dirac
Hamiltonian:
\begin{equation}
H=\beta M+\vec{\alpha}\cdot \vec{p}+\left( \beta S+V\right) ,  \label{DiracH}
\end{equation}%
where $S$ and $V$ represent the scalar potential and vector potential,
respectively. For transforming $H$ into a diagonal form, Wegner's
formulation of the SRG is adopted~\cite{Wegner94}. The initial Hamiltonian $%
H $ is transformed by the unitary operator $U(l)$ according to
\begin{equation}
H(l)=U(l)HU^{\dag }(l),\ \ H(0)=H  \label{Utransfomation}
\end{equation}%
where $l$ is a flow parameter. Differentiation of Eq. (\ref{Utransfomation})
gives the flow equation as
\begin{equation}
\frac{d}{dl}H\left( l\right) =\left[ \eta \left( l\right) ,H\left( l\right) %
\right] ,  \label{floweq}
\end{equation}%
with the generator
\begin{equation}
\eta \left( l\right) =\frac{dU\left( l\right) }{dl}U^{\dagger }\left(
l\right) =-\eta ^{\dagger }\left( l\right) .  \label{generator}
\end{equation}%
The generator $\eta (l)$ should be chosen in such a way, so that the
off-diagonal matrix elements decay. A good choice is defined by $\eta
(l)=[H_{d}(l),H(l)]$, where $H_{d}(l)$ is the diagonal part of $H(l)$\cite%
{Wegner94}. For the Dirac Hamiltonian (\ref{DiracH}), it is appropriate to
choose $\eta (l)=[\beta M,H(l)]$\cite{Guo121}. In the choice of $\eta (l)$, $%
H\left( l\right) $ can be evolved into a diagonal form in the limit $%
l\rightarrow \infty $. By using the technique in Ref. \cite{Guo121}, we have
obtained the diagonalized Dirac operator as%
\begin{equation}
H_{D}=\left(
\begin{array}{cc}
H_{P}+M & 0 \\
0 & -H_{P}^{C}-M%
\end{array}%
\right) ,  \label{diagDiracH}
\end{equation}%
where%
\begin{eqnarray}
H_{P} &=&\Sigma +\frac{p^{2}}{2M}-\frac{1}{2M^{2}}\left( Sp^{2}-\nabla
S\cdot \nabla \right)  \notag \\
&&+\frac{1}{4M^{2}}\vec{\sigma}\cdot \left[ \nabla \Delta \times \vec{p}%
\right] +\frac{1}{8M^{2}}\nabla ^{2}\Sigma  \notag \\
&&+\frac{S}{2M^{3}}\left( Sp^{2}-2\nabla S\cdot \nabla \right) -\frac{S}{%
2M^{3}}\vec{\sigma}\cdot \left[ \nabla \Delta \times \vec{p}\right]  \notag
\\
&&-\frac{1}{16M^{3}}\left[ \left( \nabla \Sigma \right) ^{2}-2\nabla \Sigma
\cdot \nabla \Delta +4S\nabla ^{2}\Sigma \right]  \notag \\
&&-\frac{p^{4}}{8M^{3}}+O\left( \frac{1}{M^{4}}\right) ,  \label{DiracHp}
\end{eqnarray}%
is an operator describing Dirac particle. Its charge-conjugation $H_{P}^{C}$
is an operator describing Dirac anti-particle. $\Sigma =V+S$ and $\Delta =V-S$
denote the combinations of the scalar potential $S$ and the vector potential
$V$. Different from Ref.~\cite{Guo121}, $H_{P}$ here is applicable for any
deformed system. From Eq. (\ref{DiracHp}), we can see that the operators
reflecting the spin-orbit interaction and the dynamic effect have been
extracted explicitly from the original Dirac Hamiltonian. To make it clear,
we decompose $H_{P}$ into five terms: $\Sigma +\frac{p^{2}}{2M}$, $-\frac{1}{%
2M^{2}}\left( Sp^{2}-\nabla S\cdot \nabla \right) +$ $\frac{S}{2M^{3}}\left(
Sp^{2}-2\nabla S\cdot \nabla \right) $,$\frac{1}{4M^{2}}\vec{\sigma}\cdot %
\left[ \nabla \Delta \times \vec{p}\right] $ $-\frac{S}{2M^{3}}\vec{\sigma}%
\cdot \left[ \nabla \Delta \times \vec{p}\right] $, $-\frac{p^{4}}{8M^{3}}$,$%
\frac{1}{8M^{2}}\nabla ^{2}\Sigma -\frac{1}{16M^{3}}\left[ \left( \nabla
\Sigma \right) ^{2}-2\nabla \Sigma \cdot \nabla \Delta +4S\nabla ^{2}\Sigma %
\right] $, which are respectively labeled as $O_{1},O_{2},\ldots ,O_{5}$. In
this decomposition, every term $O_{i}(i=1,2,\ldots ,5)$ is Hermitian. $O_{1}$
corresponds to the operator describing Dirac particle in the nonrelativistic
limit. $O_{2}$ is related to the dynamical effect. The spin-orbit
interaction is reflected in the $O_{3}$. $O_{4}$ is the relativistic
modification of kinetic energy. As $O_{i}$ is Hermitian, we can calculate
the contribution of every term to the single particle energies, which is
helpful to disclose the origin of relativistic symmetries. Especially, we can explore the deformation driven effect of the spin-orbit interaction and
dynamical term, which is interesting not only for nuclei, but also for
quantum controls and materials designs.

As $H_{P}$ is appropriate for any deformed system. As an example, we apply
it to an axially quadrupole-deformed nucleus. The corresponding potentials
are adopted as~\cite{Li10}
\begin{eqnarray}
S\left( \vec{r}\right) &=&S_{0}\left( r\right) +S_{2}\left( r\right)
P_{2}\left( \theta \right) ,  \notag \\
V\left( \vec{r}\right) &=&V_{0}\left( r\right) +V_{2}\left( r\right)
P_{2}\left( \theta \right) ,  \label{potential}
\end{eqnarray}%
where $P_{2}\left( \theta \right) =\frac{1}{2}\left( 3\cos ^{2}\theta
-1\right) $. The radial parts in Eq.(\ref{potential}) take the Woods-Saxon
form,
\begin{eqnarray}
S_{0}\left( r\right) &=&S_{\text{WS}}f(r),\text{\ }S_{2}\left( r\right)
=-\beta _{2}S_{\text{WS}}k\left( r\right) ,  \notag \\
V_{0}\left( r\right) &=&V_{\text{WS}}f(r),\text{ }V_{2}\left( r\right)
=-\beta _{2}V_{\text{WS}}k\left( r\right) ,  \label{radialpotential}
\end{eqnarray}%
with $f\left( r\right) =\frac{1}{1+\exp \left( \frac{r-R}{a}\right) }$, and $%
k\left( r\right) =r\frac{df\left( r\right) }{dr}$. Here $V_{\text{WS}}$ and $%
S_{\text{WS}}$ are, respectively, the typical depths of the scalar and
vector potentials in RMF chosen as 350 and -405 MeV, the diffuseness of the
potential $a$ is fixed as 0.67 fm, and $\beta _{2}$ is the axial deformation
parameter of the potential. The radius $R\equiv r_{0}A^{1/3}$ with $%
r_{0}=1.27$ fm. $^{154}$Dy is chosen as an example. The energy spectra of $%
H_{P}$ are calculated by expansion in harmonic oscillator basis. The
contribution of $O_{i}$ to the energy level $E_{k}$ is calculated by the formula $%
\left\langle k\right\vert O_{i}\left\vert k\right\rangle =\int \psi
_{k}^{\ast }O_{i}\psi _{k}d^{3}\vec{r}$, where $k$ marks the single particle
state considered.

In Fig.1, we show the variations of the single particle energies from all
the relativistic modifications with $\beta _{2}$ for a pair of spin doublet,
which are labeled by the corresponding spherical states with $\Omega =1/2$.
The spin-orbit term and dynamical term play a dominant role in the
relativistic modifications of the energies, while the influences from the
other terms is minor. Furthermore, the relativistic modification contributed
by the spin-orbit interaction is remarkably associated with $\beta _{2}$.
The same case also appears in the dynamical term. Over the range of
deformation under consideration, the energies contributed by the spin-orbit
term are negative for the spin aligned states and positive for the spin
unaligned states, while those by the dynamic term are always positive. The
same conclusions can be obtained for all the states with $l\neq 0$ in the
spherical notation.

Considering that the relativistic effect originates mainly from the
spin-orbit interaction and dynamical effect, it is necessary to compare
their dependencies on the deformation for the different angular momentum
states. In Fig. 2, we display the energies contributed by the spin-orbit
term and dynamical term varying with $\beta _{2}$ for the four pairs of spin
doublets with $\Omega =1/2$. The spin-orbit splittings at $\beta _{2}=0$ are
almost the most remarkable. With the increasing of deformation toward the
oblate or prolate direction, the spin-orbit splittings reduce. For all the
doublets considered, the energies contributed by the spin-orbit term is
negative for the spin aligned states and positive for the spin unaligned
states, and the spin-orbit splittings increase with the increasing angular
momentum. Different from the spin-orbit interaction, the energies
contributed by the dynamic term first decrease, then increase with the
evolution of deformation from oblate to prolate, and are more sensitive to
the deformation for the states with higher angular momentum. Similar to that
in Fig.1, the energies by the dynamical term is always positive. The same
conclusions can be obtained for the states with $\Omega >1/2$.

As the deformation influences the spin-orbit interaction and dynamical
effect, which play a critical role in the energy level structure, it is
interesting to explore the PSS and its origin for deformed nuclei. In Figs.
3 and 4, we show the variations of the energy splitting from every term with
$\beta _{2}$ for the four pairs of pseudospin doublets. The variation of
total energy splitting with $\beta _{2}$ is dominated by the three parts:
the nonrelativistic term, the spin-orbit term, and the dynamic term. The
influences from the relativistic modification of kinetic energy and the
other term are almost negligible. The conclusion is consistent with that for
the spherical nuclei~\cite{Guo121}. Over the range of $\beta _{2}$ here, the
energy splitting from the nonrelativistic term is the most remarkable. The
relativistic PSS is significantly improved, which comes mainly from the
spin-orbit interaction and dynamical effect. Compared with the dynamical effect, the spin-orbit interaction is more sensitive to $\beta _{2}$ in the oblate side. The PSS becoming worse with the increasing of $|\beta _{2}|$ is mainly due to the weaker spin-orbit interaction. In the prolate side, the PSS becomes worse with the increasing of $\beta_{2}$ for the doublets $(1/2[420],3/2[422])$ and $(1/2[530],3/2[532])$, which is attributed to a combination of the weaker spin-orbit improvement and the stronger dynamical breaking (or the weaker dynamical improvement). However for the doublets $(5/2[402],7/2[404])$ and $(5/2[512],7/2[514])$, the spin-orbit improvement becomes stronger and the dynamical breaking becomes weaker (or the dynamical improvement becomes stronger) with $\beta_{2}$, which result in the PSS becoming better. These have explained the reason why the PSS becomes worse for the more bound energy levels and better for the energy levels closer to the continuum (to see Fig.5) for most of the heavy nuclei holding prolate shape.

In order to better grasp the PSS in deformed nuclei, the single particle
energies for all the pseudospin doublets are plotted against the deformation
$\beta _{2}$ ranging from -0.3 to 0.5 in Fig. 5, where the pseudospin
doublets are labeled with the asymptotic Nilsson quantum numbers $\Omega
\lbrack N,n_{3},\Lambda ]$. For zero deformation $\beta _{2}=0$, the orbits
are indicated by the corresponding spherical states. The figure reveals the
following: (i) The energy difference between the pseudospin unaligned and
aligned states always remains positive over the range of deformation
considered here. (ii) The energy splittings between the pseudospin partners
are more sensitive to $\beta _{2}$ for the oblate side than that for the
prolate side. (iii) The energy splitting between the pseudospin partners is
smaller for the valence orbits and for the partners just below the Fermi
surface. The systematics has been explained well in the preceding analysis.

In summary, we apply the similarity renormalization group to transform a
general Dirac Hamiltonian into diagonal form. The diagonal Dirac operator
consists of the nonrelativistic term, the spin-orbit term, the dynamic term,
and the relativistic modification of kinetic energy, and the other term,
which are very useful to explore the symmetries hidden in the Dirac
Hamiltonian. As an example, we have checked the relativistic symmetries for
an axially deformed nucleus by comparing the contributions of every term to
the single particle energies and their correlations with the deformation. It
is shown that the spin-orbit interaction and dynamical effect play the
key roles in the PSS. Their contributions to the pseudospin
energy splitting are correlated with the deformation of the potential and
the quantum numbers of the state. Over the range of the deformation considered here, the spin-orbit interaction always improves the PSS, while the dynamical effect relates to the deformation and the particular state. For the deeply bound energy levels, the
contribution of the dynamical term is a breaking of the PSS, while for the energy
levels near to the continuum, the contribution of dynamical term becomes an
improvement to the PSS. Compared with the dynamical effect, the dependence of the spin-orbit interaction on the deformation is more sensitive, which dominates the change of PSS in the oblate side. In the prolate side, with the development of energies with deformation toward the continuum, that the PSS becomes better is due to the stronger spin-orbit improvement and the weaker dynamical breaking (or the stronger dynamical improvement). The cause of better PSS for the levels closer to the continuum has been disclosed and the systematics of PSS associated with the deformation has been clarified.

This work was partly supported by the National Natural Science Foundation of
China under Grants No. 11175001, and No. 11205004; the Program for New
Century Excellent Talents in University of China under Grant No.
NCET-05-0558; the Excellent Talents Cultivation Foundation of Anhui Province
under Grant No. 2007Z018; the Natural Science Foundation of Anhui Province
under Grant No. 11040606M07; and the 211 Project of Anhui University.

\newpage
\begin{figure}[h!]
\includegraphics[width=9.5cm]{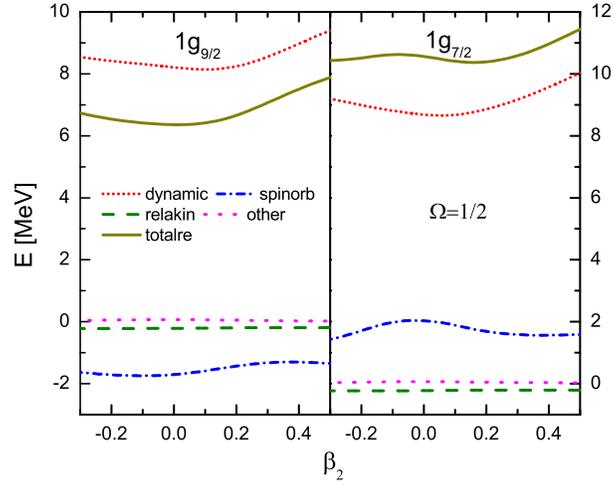}
\caption{(Color online) Comparisons of the contributions of all the
relativistic modifications to the single particle energies and their
correlations with the deformation parameter $\protect\beta_2$ for a pair of
spin doublet, which are indicated by the corresponding spherical states with
$\Omega=1/2$. The `dynamic, spinorb, relakin, and other` denote the
dynamical term, the spin-orbit term, the relativistic modification of
kinetic energy, and the other term, respectively. For guiding eyes, a sum of
all the relativistic modifications is marked as `totalre`.}
\end{figure}

\begin{figure}[h!]
\includegraphics[width=9.5cm]{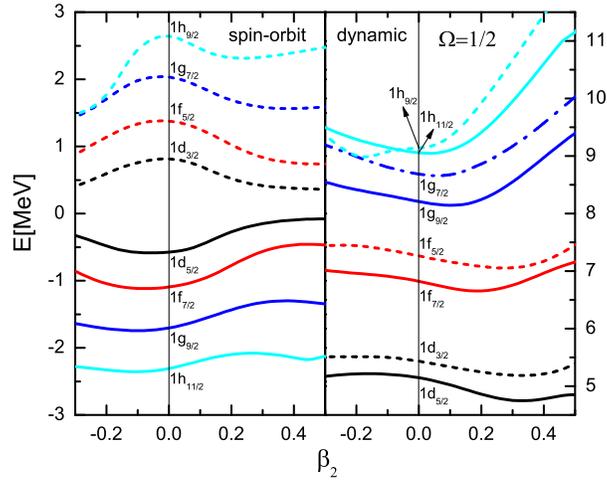}
\caption{(Color online) The contributions of the spin-orbit term and the
dynamical term to the single particle energies and their correlations with
the deformation parameter $\protect\beta_2$ for four pairs of spin doublets
with $\Omega=1/2$. }
\end{figure}

\begin{figure}[h!]
\includegraphics[width=9.5cm]{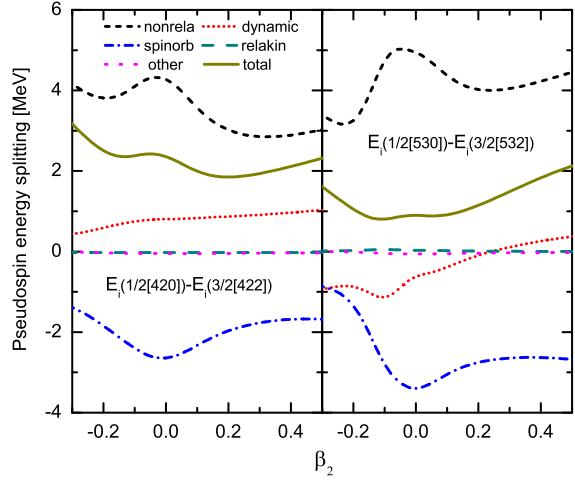}
\caption{(Color online) Comparisons of the contributions of all the terms in
$H_P$ to the pseudospin energy splittings and their correlations with the
deformation parameter $\protect\beta_2$ for the doublets $%
(1/2[420],3/2[422]) $ and $(1/2[530],3/2[532])$. The `nonrela, dynamic,
spinorb, relakin, and other` denote the nonrelativistic part, the dynamical
term, the spin-orbit term, the relativistic modification of kinetic energy,
and the other term, respectively. For guiding eyes, the total pseudospin
energy splitting is marked as `total`.}
\end{figure}

\begin{figure}[h!]
\includegraphics[width=9.5cm]{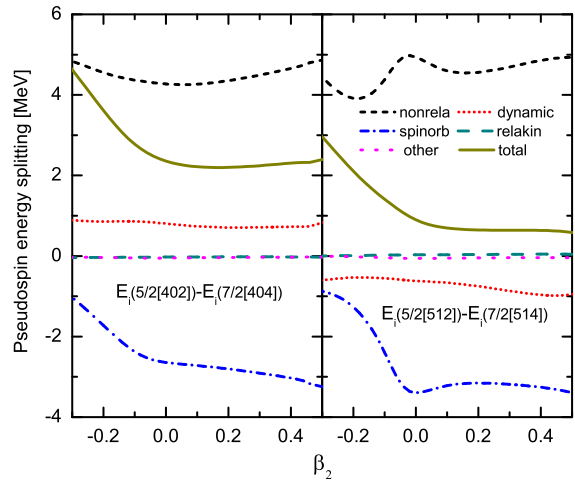}
\caption{(Color online) The same as Fig.3, but for the doublets $%
(5/2[402],7/2[404])$ and $(5/2[512],7/2[514])$.}
\end{figure}

\begin{figure}[ht]
\includegraphics[width=8cm]{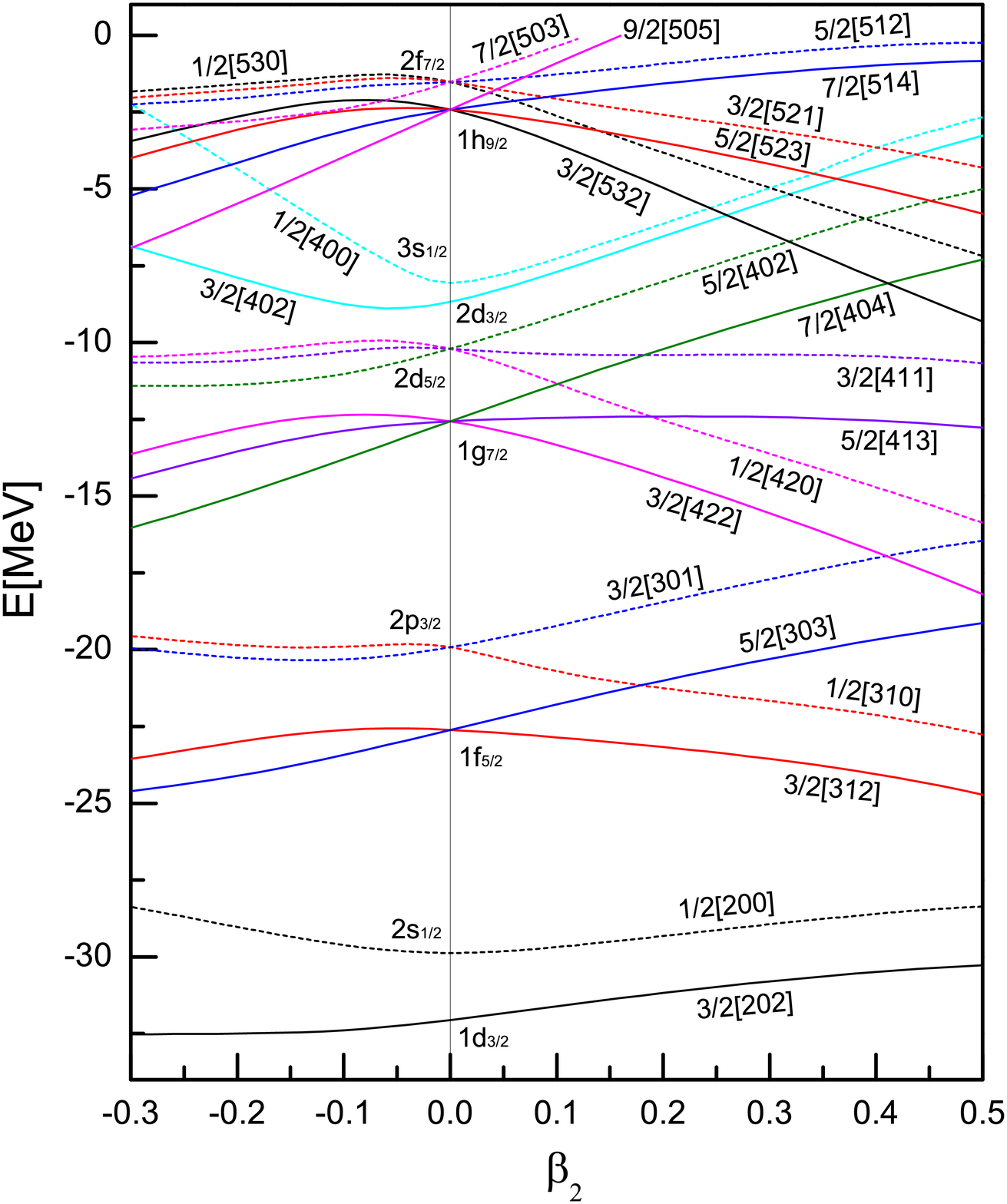}
\caption{(Color online) The single particle levels for all the pseudospin
doublets in the nucleus $^{154}$Dy as a function of the quadrupole
deformation parameter $\protect\beta_2$.}
\end{figure}


\begin{thebibliography}{99}
\bibitem{Haxel49} O. Haxel, J. H. D. Jensen, and H. E. Suess, Phys. Rev.
\textbf{75}, 1766 (1949).

\bibitem{Mayer49} M. Goeppert-Mayer, Phys. Rev. \textbf{75}, 1969 (1949).

\bibitem{Hecht69} K. T. Hecht and A. Adler, Nucl. Phys. A \textbf{137}, 129
(1969).

\bibitem{Arima69} A. Arima, M. Harvey, and K. Shimizu, Phys. Lett. B \textbf{%
30}, 517 (1969).

\bibitem{Bohr82} A. Bohr, I. Hamamoto, and B. R. Mottelson, Phys. Scr.
\textbf{26}, 267 (1982).













\bibitem{Ginoc97} J. N. Ginocchio, Phys. Rev. Lett. \textbf{78}, 436 (1997).

\bibitem{Meng98} J. Meng, K. Sugawara-Tanabe, S. Yamaji, P. Ring, and A.
Arima, Phys. Rev. C \textbf{58}, R628 (1998).





























\bibitem{Alber01} P. Alberto, M. Fiolhais, M. Malheiro, A. Delfino, and M.
Chiapparini, Phys. Rev. Lett. \textbf{86}, 5015 (2001).


\bibitem{Lisboa10} R. Lisboa, M. Malheiro, P. Alberto, M. Fiolhais, and A.
S. de Castro, Phys. Rev. C \textbf{81}, 064324 (2010).




\bibitem{Zhou03} S. G. Zhou, J. Meng, and P. Ring, Phys. Rev. Lett. \textbf{%
91}, 262501 (2003).

\bibitem{Leviatan04} A. Leviatan, Phys. Rev. Lett. \textbf{92}, 202501
(2004).

\bibitem{Typel08} S. Typel, Nucl. Phys. A \textbf{806}, 156 (2008).

\bibitem{Leviatan09} A. Leviatan, Phys. Rev. Lett. \textbf{103}, 042502
(2009).

\bibitem{Liang11} H. Z. Liang, P. W. Zhao, Y. Zhang, J. Meng, and N. V.
Giai, Phys. Rev. C \textbf{83}, 041301(R) (2011).

\bibitem{Gonoc11} J. N. Ginocchio, J. Phys. Conf. Ser. \textbf{267}, 012037
(2011).

\bibitem{Guo05} J. Y. Guo, R. D. Wang, and X. Z. Fang, Phys. Rev. C \textbf{%
72}, 054319 (2005).

\bibitem{Lu12} B. N. Lu, E. G. Zhao, and S. G. Zhou, Phys. Rev. Lett.
\textbf{109}, 072501 (2012).


\bibitem{Ginoc05PR} J.N. Ginocchio, Phys. Rep. \textbf{414}, 165 (2005).

\bibitem{Liang13} H. Z. Liang, S. H. Shen, P. W. Zhao, J. Meng, Phys. Rev. C
\textbf{87}, 014334 (2013).

\bibitem{Guo121} J. Y. Guo, Phys. Rev. C \textbf{85}, 021302(R) (2012)

\bibitem{Liang132} S. H. Shen, H. Z. Liang, P. W. Zhao, S. Q. Zhang, J.
Meng, arXiv:1308.1143v1 [nucl-th] (2013).

\bibitem{Guo122} S. W. Chen and J. Y. Guo, Phys. Rev. C \textbf{85}, 054312
(2012).

\bibitem{Guo13} D. P. Li, S. W. Chen, and J. Y. Guo, Phys. Rev. C \textbf{87}%
, 044311 (2013).

\bibitem{Wegner94} F. Wegner, Annalen der Physik (Leipzig) \textbf{506}, 77
(1994).

\bibitem{Li10} Z. P. Li, J. Meng, Y. Zhang, S. G. Zhou, and L. N. Savushkin,
Phys. Rev. C \textbf{81}, 034311 (2010).

\end{thebibliography}
\end{document}